\documentstyle[12pt]{article}
\begin{document}
\title{\bf  Nuclear transparency and the onset of strong absorption regime in the $^{12}C+$$^{24}Mg$ system}
\vskip 2.0cm 
\author{
R. Lichtenth\"aler, A. L\'epine-Szily and  M. S. Hussein\\
Departamento de F\'{\i}sica Nuclear, Instituto de F\'{\i}sica, Universidade de S\~ao Paulo\\ 
C.P.66318, 
05315-970 S\~ao Paulo, SP, Brazil\\}
\maketitle

\begin{abstract}

The elastic scattering of $^{12}C+^{24}Mg$ has been studied by means of a phase-shift analysis of 21 angular distributions ranging from $E_{lab}=16$ MeV up to $E_{lab}=40$ MeV. A tri-dimensional plot of the reflection coefficient of the S-matrix as a function of the angular momentum and the energy shows a  well defined region of energy, which separates two regimes: strong absorption for higher energies and
the so called "anomalous transparency regime", recently observed in this 
system at low energies. The Argand diagrams of the S-matrix in  angular momentum space also present very contrasting behaviours in the two regions with very rapidly varying phases in the low energy region, which we associate with a parity dependent term  in the S-matrix directly related to significant coupling to the elastic transfer of a $^{12}C$ nucleus.

\end{abstract}

An anomalous transparency has been observed\cite{al1} in the $^{12}C+^{24}Mg$ system
at energies around and  slightly above the Coulomb barrier ($E \leq 1.5E_b$).
The angular distributions of the elastic scattering are characterized by oscillations and require a very small imaginary part in the optical  potential in order to be fitted. These oscillations are present in practically the whole angular distribution, even at forward and intermediate angles, in contrast to the well-known phenomenon of Anomalous Large Angle 
Scattering (ALAS), which is observed at slightly higher energies ($E \ge 2.5E_b$)
and is characterized by oscillating angular distributions and excitation functions at large angles\cite{ba}. The oscillations observed in the angular distributions of the systems which exhibit ALAS phenomenon can be reproduced by a background strong absorption S-matrix plus a localized perturbation normally around the grazing angular momentum. This is an indication that a peripheral process is responsible for the oscillations observed at large angles. The coupling of alpha transfer channels to the elastic scattering can be the origin of this perturbation in the S-matrix\cite{ru3,ru1,ru2,al2}.
 On the other hand, the anomalous transparency at low energies shows a sensitivity of the elastic scattering to the nuclear interior,  whose signatures in the S-matrix are a very rapid variation of the phases and strong oscillations in the reflection coefficients, in contrast to a typical strong absorption 
S-matrix, which is a  smooth function of the angular momentum. 

In this paper we present the results of an extensive phase-shift analysis of the 21 available angular distributions of the   $^{12}C+^{24}Mg$ elastic scattering\cite{ru1,al2,sci,me} which show the clear transition that occurs  between the two regimes. 
The phase-shift analysis was performed using a method we recently developed 
\cite{van},  which allows the determination of the elastic S-matrix elements and their associated  errors, using a chi-square
minimization algorithm specially developed for the scattering problem. 
The method has proven to be quite versatile at  low energies, where the convergence is achieved very rapidly. 

There are two important aspects in the phase-shift analysis: the starting S-matrix for the search and the maximum number of partial waves used in the 
search $(lmax)$. As a starting point for the analysis, we used the S-matrix generated by the optical model potential named X3 
in ref. \cite{ru1}. This potential reproduces well the forward angle region and
the mean slope of the angular distributions in the range of energy studied, 
but not the oscillations. 

 For $l \geq lmax$ we use the S-matrix elements calculated by the X3 optical potential. For $l \l lmax$ the S-matrix elements are obtained by search.
The maximum angular momentum in the analysis was choosen by the condition $|S_{lmax}| \geq .93$.
This analysis is more reliable, as far as ambiguities are concerned due to
the fact that the errors of the S-elements are also calculated. We observed that when the error bars are small, the final S-matrix is practically independent of the starting point of the search and of the maximum number of partial waves  used in the analysis. On the other hand, if the error bars are big a large ambiguity in the final S-matrix is observed. The precision of the final S-matrices is related to the quality and completeness of the experimental angular distributions. 

The result of this analysis is presented in figure 1 in a tri-dimensional plot of the reflection coefficients ($|S_l(E)|$) as a function of the laboratory 
energy and of the angular momentum. We can observe the clear transition between the two regimes occuring at laboratory energies around 28 MeV. The Coulomb barrier for this system is 
$E_{lab}=18.5$ MeV, calculated using the best fit optical potential of 
reference \cite{al1}.
The reflection coefficients for energies above 28MeV  show a relatively smooth transition
from  unity, at high angular momentum values, to very low values at low partial waves. This is a typical strong absorption behaviour with a well defined value of the grazing angular momentum. The very small perturbations observed in  
this S-matrix at low l-values are responsible for the ALAS phenomenon.  Below 28MeV the situation is completely different,
with strongly oscillating reflection coefficients as a function of the energy and angular momentum. In this situation, the grazing trajectories are not well defined in the sense that many different values of angular momentum can be assigned to the condition $|S_l|=0.5$. 

In figures 2a,b,c,d we plot the reflections coefficients(left) and  Argand 
 diagrams(right) $(S_r X S_i)$ of the S-matrices as a function of the angular momentum at four laboratory energies: 40, 31.2, 23 and 20MeV. At the two higher energies we observe a typical strong absorption behaviour either in the reflection coefficients or in the Argand
diagrams which start at $S=1$ for high angular momenta and go smoothly to zero for low angular momenta. The Argand diagram is restricted to  the first quarter of the complex plane $S_R \geq 0$ and $S_I \geq 0$ showing a small positive nuclear phase corresponding to an attractive nuclear potential. 
At lower
 energies, on the other hand,  one can observe  loops in the Argand diagrams showing a rapidly increasing real phase for decreasing angular momenta. 
The presence of these loops traversed in the anti-clockwise sense,
 as l decreases, 
can be due to a parity dependent term in the S-matrix, as we show qualitatively 
below.

In fig.2e we plot the S-matrix obtained by the simple parametrization:
\begin{equation}
S=S_0+(-1)^lS_{et}
\end{equation}
where $S_0$ is a background S-matrix taken to be  of the Ericsson form:
\begin{equation}
S_0=\frac{1}{1+exp(\frac{(lg_0-l)}{\Delta_0}-i \alpha_0)}
\end{equation} 

and $S_{et}$ is a parity dependent term given by:

\begin{equation}
S_{et}=A\frac{exp(\frac{lg-l}{\Delta} -i \alpha)}{(1+exp(\frac{lg-l}{\Delta}-i \alpha))^2}
\end{equation}

The values of the parameters are given:$lg_0=7$, $\Delta_0=1.0$, $\alpha_0=0.7$ 
$lg=4$, $\Delta=1.0$, $\alpha=-0.7$, and $A=2.5$. 
The loops  in the corresponding Argand diagrams in fig. 2e(right), are introduced by the term $S_{et}$. In order to have the correct sense, loop traversed in anti-clockwise sense as l decreases, it is necessary to use a negative nuclear phase parameter 
$\alpha$.                
If $\alpha$ is positive, the sense of traversing the loops is opposite to the observed in the experimental Argand diagrams. As a negative nuclear phase corresponds to a repulsive  nuclear potential, we conclude  that the interaction responsible for the parity dependence is repulsive. This is in agreement with the analysis presented in ref. \cite{al3} where we found a positive form factor. 

The elastic transfer of a $^{12}C$ between the $^{24}Mg$ target and the $^{12}C$ projectile can be the physical origin of this parity dependence in the S-matrix as suggested in \cite{al3}. 

Another possible explanation for the loops observed in the Argand diagrams in l-space is the presence of Regge poles, which would come closer to the real axis in the low energy region, where transparency comes in. The presence of Regge poles would be an indication of a resonant regime at these energies, probably due to quasi-molecular states in the $^{36}Ar$ composite system. In order to test this hypothesis, 
 we plotted Argand diagrams of the S-matrices as a function of  energy for fixed angular momentum. The Argand diagrams in the complex energy plane show an apparently chaotic behaviour and no loops were observed for any angular 
momentum involved in the scattering. This indicates that the resonances, if present,  must have widths smaller than 330 KeV(C.M.) which is the energy step of the measurements at low energies. For this reason we discard the hypothesis
of  molecular resonances of widths greater than 330 KeV.

An interesting phenomenon observed in our analysis, is the sudden onset of strong absorption at energies $\approx 8$ MeV(c.m) above the Coulomb barrier.
 This could  be associated with a threshold effect related to the opening of the $3- \alpha$ cluster (Hoyle) resonance channel in $^{12}C$.

Recently we have presented evidence, based on optical model analysis, 
in support of strong coupling of the elastic  $^{12}C+$$^{24}Mg$ channel to the elastic transfer channel $^{24}Mg+$$^{12}C$ near barrier energies which strongly influences the angular distributions. 
Further evidence in support of 
a strong two $^{12}C$ clustering effect in the ground state of $^{24}Mg$
is presented here. 
This finding has important implications on the two $^{12}C$ cluster nature of the ground state of $^{24}Mg$
and may potentially be significant in nuclear astrophysics. 

\noindent
{\bf Acknowledgements}

This work has been supported partially by FAPESP and CNPq.

\newpage
\section{Figure Captions}

Figure 1. Plot of the reflection coefficients as a function of the laboratory energy and angular momentum.           

\vskip 1cm

\noindent
Figure 2(left) The reflection coefficients for 20MeV(a), 23MeV(b), 
31.2MeV(c), 40MeV(d) and for the S-matrix of equation 1(e). The dashed line 
in fig2(e) is the background reflection coefficients ($|S_0|$, eq.1). 
See text for details. 

\vskip 1cm
\noindent
Figure 2(right) Argand diagrams of the S-matrix as a function of the angular momentum for the same energies and for the S-matrix of equation 1. 
See text for details.

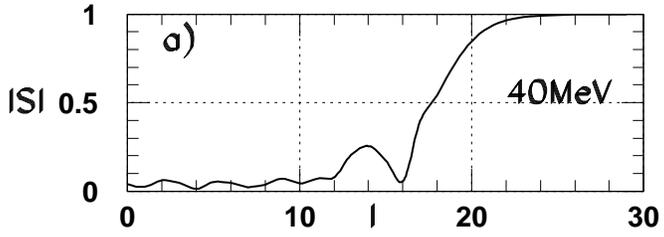
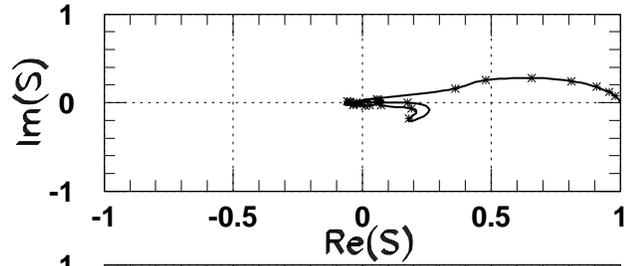
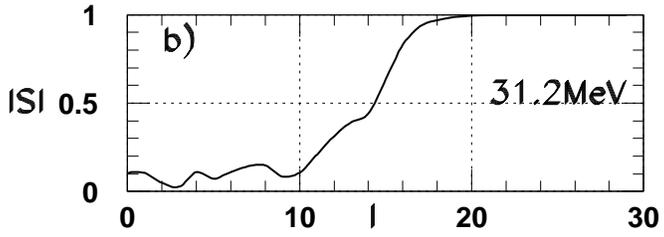
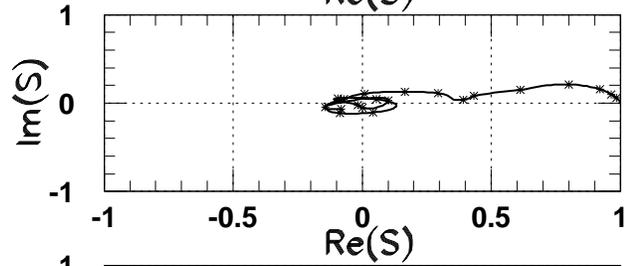
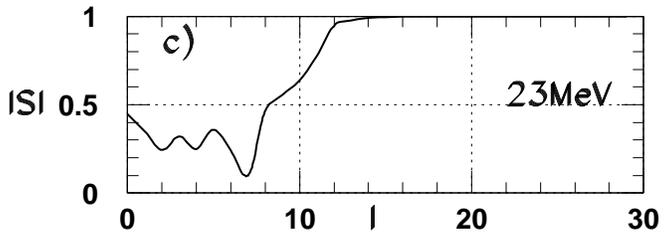
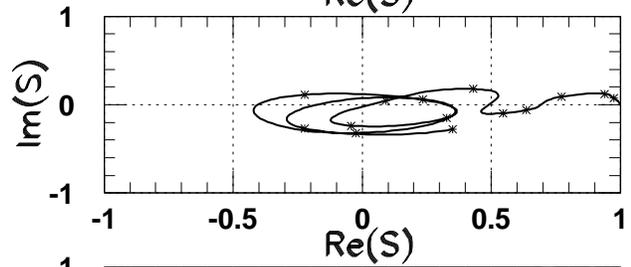
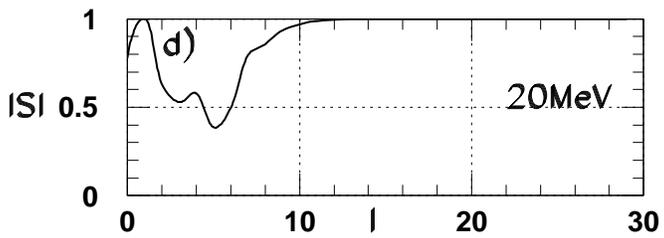
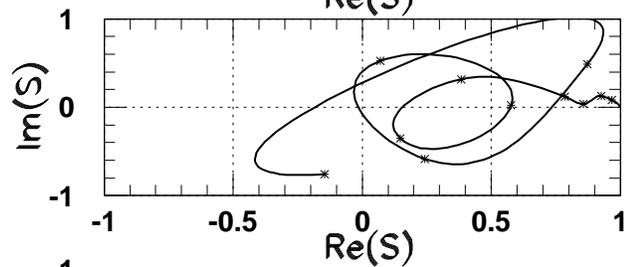
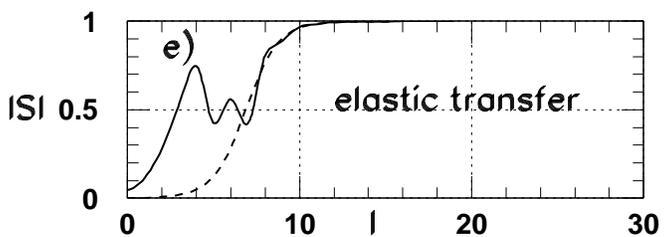
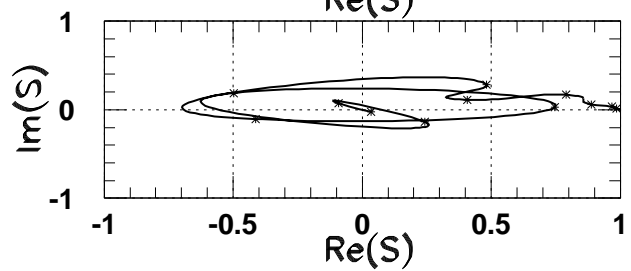
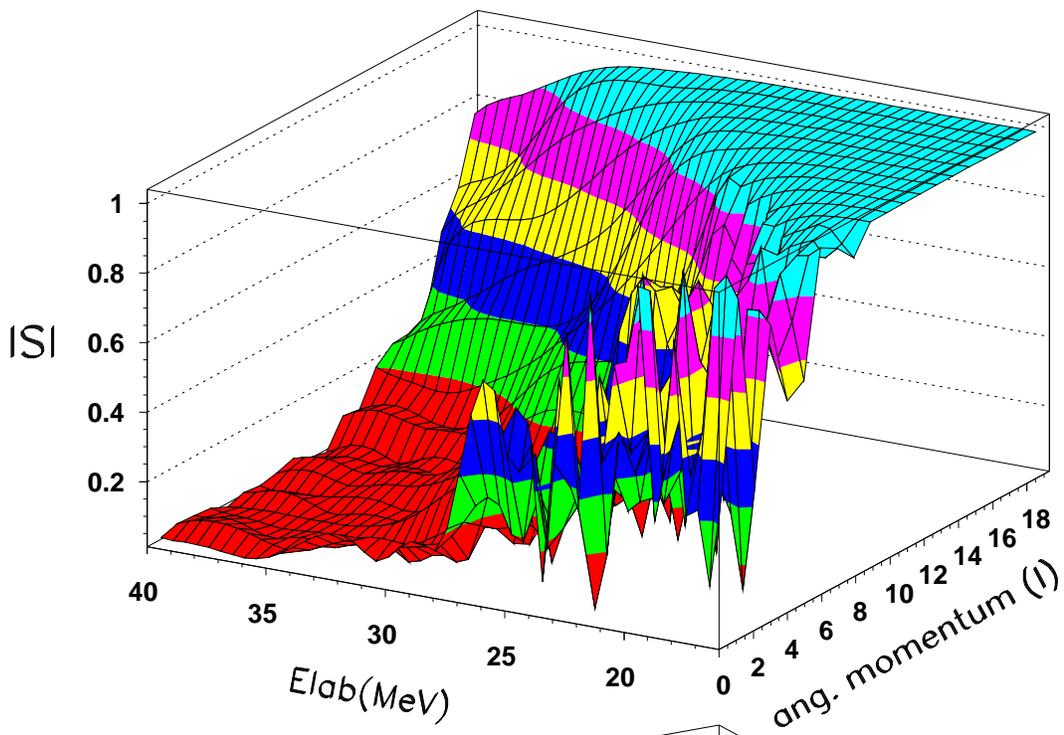
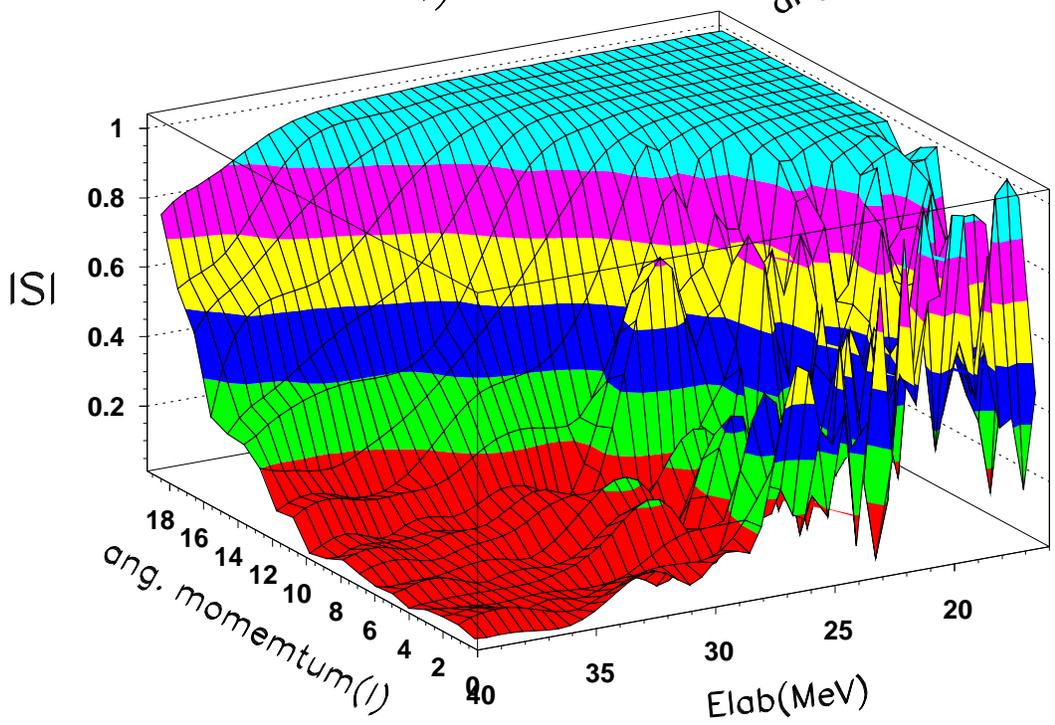
\end{document}